\documentclass[twocolumn,aps,prl,groupedaddress]{revtex4-1}

\usepackage{graphicx}
\usepackage{dcolumn}
\usepackage{bm}
\usepackage{amsmath,amssymb,amsfonts}
\usepackage{color}

\begin{document}

\title{New quantum critical points of $j=\frac{3}{2}$ Dirac electrons in antiperovskite topological crystalline insulators}
\author{Hiroki Isobe}
\author{Liang Fu}
\affiliation{Department of Physics, Massachusetts Institute of Technology,
Cambridge, Massachusetts 02139, USA}

\begin{abstract}
We study the effect of the long-range Coulomb interaction in $j=3/2$ Dirac electrons in cubic crystals with the $O_h$ symmetry, which serves as an effective model for antiperovskite topological crystalline insulators.  The renormalization group analysis reveals three fixed points that are Lorentz invariant, rotationally invariant, and $O_h$ invariant.  Among them, the Lorentz- and $O_h$-invariant fixed points are stable in the low-energy limit while the rotationally invariant fixed point is unstable. The existence of a stable $O_h$-invariant fixed point of Dirac fermions with finite velocity anisotropy presents an interesting counterexample to emergent Lorentz invariance in solids.
\end{abstract}


\maketitle

The discovery of Dirac electrons (broadly defined) in solids has opened up a variety of new topics in physics for a decade. Examples of Dirac materials include graphene \cite{graphene}, topological insulators \cite{ti1,ti2}, and Dirac/Weyl semimetals \cite{weyl1,weyl2}.  The important feature of massless Dirac fermions is the linear energy dispersion crossing at a point, which makes the theory scale invariant. Still there is a difference from the Dirac theory in high-energy physics; in solids, the speed of electrons $v$ is smaller than the speed of light $c$ and hence the Lorentz invariance is broken when electron-photon interaction is present. Also, the velocity of Dirac electrons can differ along different directions in a crystal.

Electron interactions can modify the Dirac dispersion. When the Fermi level lies at the Dirac point, the Coulomb interaction is unscreened and hence long ranged. It enhances the speed of electrons $v$ logarithmically, both in two and three dimensions \cite{mishchenko, son, kotov, two-loop, das sarma, vishwanath, 3d_strong}. One may think that $v$ has a logarithmic divergence in the low-energy limit, but the relativistic effect, namely, the coupling to the electromagnetic field, makes it converge to the speed of light $c$ \cite{guinea, rg1, herbut}.  This is an example of emergent Lorentz invariance as a low-energy phenomenon \cite{nielsen}. It is also true for two-dimensional anisotropic Weyl semimetals with linear but tilted energy dispersion \cite{rg_tilt}.

Qualitatively different results appear for generalized Dirac electrons whose energy dispersion deviates from linearity. For example, when two Weyl cones move and merge in the Brillouin zone, the energy dispersion will be quadratic along the merging direction. In such cases, stable fixed points are anisotropic in three dimensions \cite{rg_3d} and non-Fermi liquid or marginal Fermi liquid in two dimensions \cite{rg_2d, rg_2dcrit}.
A non-Fermi-liquid state is also theoretically discovered in the Luttinger Hamiltonian with a quadratic band touching in three dimensions \cite{3d_quad}.
Other nontrivial fixed points are found in three-dimensional double-Weyl semimetals \cite{3d_dbweyl1, 3d_dbweyl2} and nodal-ring semimetals \cite{3d_nodal}.

Recently, a new type of Dirac electrons has been theorized \cite{hsieh2} in antiperovskite materials $A_3BX$ with $A=(\text{Sr, La, Ca})$, $B= (\text{Sn, Pb})$ and $X=(\text{O, N, C})$. These materials are predicted to be in or very close to a topological crystalline insulator (TCI) phase \cite{AndoFu}. This TCI phase was previously discovered in IV-VI semiconductors Sn$_{1-x}$Pb$_x$(Te,Se) \cite{hsieh, ando, poland, hasan} and has stimulated wide interest. In both classes of materials, the nontrivial topology is protected by mirror symmetry and results from band inversion described by the sign change of the Dirac mass. However, unlike IV-VI semiconductors, antiperovskites have a fundamental band gap located at $\Gamma$, where both the conduction and valence bands are four-fold degenerate consisting of $j=3/2$ quartets. The band structure near $\Gamma$ is well described by a first-order eight-component $k\cdot p$ Hamiltonian \cite{hsieh2}, which is a high-spin generalization of the Dirac equation for spin-1/2 particles.

In this paper, we report quantum critical points of such $j=3/2$ Dirac electrons in cubic crystals with the $O_h$ symmetry. The system has linearly dispersing energy bands in all directions, with anisotropic velocity parameters reflecting the $O_h$ symmetry. Based on renormalization group (RG) analysis, we find in the presence of Coulomb interaction, $j=3/2$ Dirac electrons exhibit three fixed points that are Lorentz invariant, rotationally invariant, and $O_h$ invariant, respectively. The rotationally invariant fixed point is unstable and flows to the Lorentz- and $O_h$-invariant fixed points that are stable.
The existence of the stable $O_h$ fixed point, with a finite velocity anisotropy, is rather unusual and contrasts with previously known Dirac systems with linearly dispersing energy bands which all exhibit emergent Lorentz invariance.

\textit{Model}.
The effective Hamiltonian for $j=3/2$ Dirac octets is
\begin{equation}
H(\bm{k}) = m\tau_z + v_1 \tau_x \bm{k}\cdot \bm{J} + v_2 \tau_x \bm{k}\cdot \tilde{\bm{J}},
\end{equation}
where $\bm{J}$ is a set of spin-3/2 matrices and $\tilde{\bm{J}}$ is a set of $4\times 4$ matrices that transforms as a vector under the cubic point group $O_h$.
$\tilde{\bm{J}}$ is also written as a linear combination of $\bm{J}$ and $\bm{J}^3$.
We note that $\bm{k}\cdot\bm{J}$ respects the rotational symmetry, while $\bm{k}\cdot\tilde{\bm{J}}$ does not.
Since $\bm{J}$ are the generators of rotation, which is continuous symmetry, their commutation relations are in closed form
\begin{equation}
[ J^i, J^j ] = i\epsilon^{ijk} J^k,
\end{equation}
where $i,j,k$ correspond to three-dimensional coordinates $x,y,z$.
In contrast, $\tilde{\bm{J}}$ satisfies
\begin{equation}
[ \tilde{J}^i, \tilde{J}^j ] = i\epsilon^{ijk} \left( \tilde{J}^k -\frac{3}{2} J^k \right),
\end{equation}
which is not closed.

The sign of the mass parameter $m$ controls the topological phase transition. We consider the quantum critical point $m=0$, where the band gap closes. Then the Hamiltonian becomes
\begin{equation}
\label{eq:ham1}
H(\bm{k}) = v_1 \bm{k}\cdot \bm{J} + v_2 \bm{k}\cdot \tilde{\bm{J}},
\end{equation}
after diagonalizing $\tau_x$. Here we assume zero chemical potential.
It is convenient to write the Hamiltonian using the following matrices:
\begin{equation}
\bm{\gamma}_d = \frac{2}{5} (\bm{J} - 2\tilde{\bm{J}}), \quad
\bm{\gamma}_s = \frac{2}{5} (2\bm{J} + \tilde{\bm{J}}),
\end{equation}
which satisfy
$\text{tr} (\gamma_d^i \gamma_d^j) = \text{tr} (\gamma_s^i \gamma_s^j) = 4\delta^{ij}$ and
$\text{tr} (\gamma_d^i) = \text{tr} (\gamma_s^i) = \text{tr} (\gamma_d^i \gamma_s^j) = 0$.
Then the Hamiltonian is rewritten as \cite{hsieh2}
\begin{equation}
\label{eq:ham2}
H(\bm{k}) = v_d \bm{k}\cdot\bm{\gamma}_d + v_s \bm{k}\cdot\bm{\gamma}_s,
\end{equation}
where the two velocity parameters are defined by
$v_d = v_1/2-v_2$ and $v_s = v_1 + v_2/2$.

The $4 \times 4$ matrices $\gamma_d^i$ satisfy the anticommutation relation
\begin{equation}
\{ \gamma_d^i, \gamma_d^j\} = 2\delta^{ij},
\end{equation}
which indicates the Hamiltonian reduces to two copies of Weyl Hamiltonians when $v_s = 0$.
It means that the present model holds the Lorentz symmetry at $v_s = 0$.
$\gamma_d^i$ and $\gamma_s^i$ follow the commutation relations
\begin{gather}
[ \gamma_d^i, \gamma_d^j ] = -2i \epsilon^{ijk} \gamma_d^k, \quad
[ \gamma_s^i, \gamma_s^j ] = i \epsilon^{ijk} \gamma_d^k, \notag \\
[ \gamma_d^i, \gamma_s^j ] + [ \gamma_s^i, \gamma_d^j] = 2i \epsilon^{ijk} \gamma_s^k,
\end{gather}
where the first equality shows that $\gamma_d^i$ are the generators of $SU(2)$ algebra.

We introduce the long-range Coulomb interaction
\begin{equation}
V(q) = \frac{e^2}{\varepsilon q^2}
\end{equation}
as a perturbation to the system.
When the Fermi energy is zero, the density of states vanishes at the Fermi level, and hence the Coulomb interaction is not screened and long ranged.

\begin{figure}
\centering
\includegraphics[width=0.8\hsize]{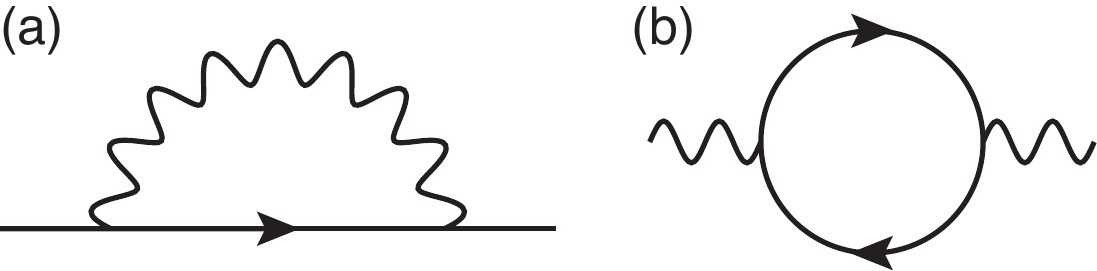}
\caption{
Radiative corrections at one-loop order: (a) self-energy and (b) polarization. Solid lines and wavy lines represent the electron propagator and the Coulomb interaction, respectively.
}
\label{fig:diagram}
\end{figure}

\textit{Renormalization group analysis}.
We consider the effect of the long-range Coulomb interaction by perturbative RG analysis.
In the following analysis, we employ the Euclidean action and calculate the radiative corrections to one-loop order (Fig.~\ref{fig:diagram}). Here the noninteracting Green's function is given by $G_0(\bm{k},i\omega) = [i\omega -H(\bm{k})]^{-1}$.

\begin{figure*}
\centering
\includegraphics[width=\hsize]{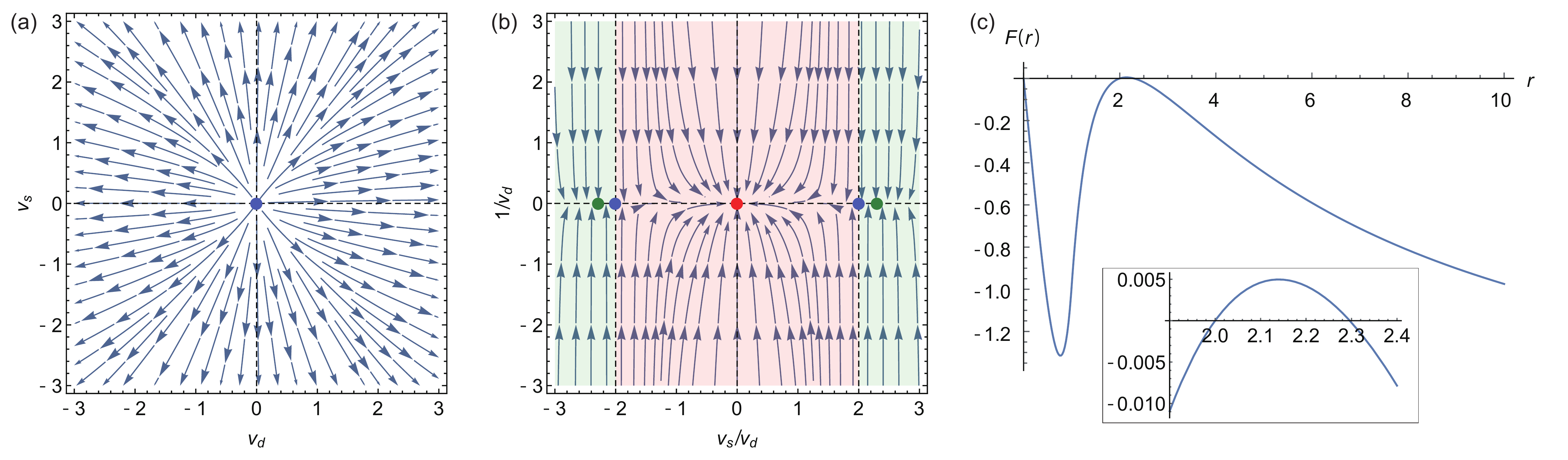}
\caption{
RG flows and fixed points. (a) RG flow of the velocities $v_d$ and $v_s$. There is an unstable fixed point (blue) at $v_d = v_s = 0$, and both $v_d$ and $v_s$ become larger as one goes to low energies.
(b) RG flow of the ratio $r=v_s/v_d$. Though both $v_d$ and $v_s$ diverge in the low-energy limit, the ratio $r$ could be finite. There are stable fixed points at $r=0$ (red) and $r=\pm r_s$ ($r_s\approx 2.296$) (green), and unstable fixed points at $r = \pm 2$ (blue).
Any value of $r$ in the red region $|r| < 2$ flows to the Lorentz-invariant fixed point at $r=0$, and $r$ in the green regions $|r| > 2$ flows to the fixed points at $r=\pm r_s$.
(c) Function $F(r)$ that determines the fixed points of the ratio $r$ [see Eq.~\eqref{eq:beta_r}].
The function $F(r)$ is an odd function of $r$. We can find zeros at $r=0, 2,$ and $r_s$, and the sign of $F(r)$ determines the stability around the zeros.
}
\label{fig:flow}
\end{figure*}

First, we calculate the one-loop self-energy $\Sigma(\bm{p}, i\omega)$ [Fig.~\ref{fig:diagram}(a)], which is given by
\begin{align}
\Sigma (\bm{p}, i\omega) &= - \int_{\bm{k},\omega'}' G_0 (\bm{k}, i\omega') V(|\bm{k}-\bm{p}|) \notag \\
& = -\frac{e^2}{\varepsilon} \int_{\bm{k},\omega'}' G_0 (\bm{k}, i\omega') \frac{2\bm{k}\cdot\bm{p}}{k^4} + O(p^2).
\end{align}
The integral $\int_{\bm{k},\omega'}'$ stands for $\int\frac{d\omega'}{2\pi} \int'\frac{d^3k}{(2\pi)^3}$, where $\int' dk$ means a momentum integration over the shell $( \Lambda e^{-l}, \Lambda]$. This momentum shell procedure regularizes a logarithmic divergence, and it gives the renormalization of the velocity parameters.
The self-energy can be decomposed as
\begin{equation}
\Sigma(\bm{p}, i\omega ) = \Sigma_0 i\omega + \Sigma_d \bm{p} \cdot \bm{\gamma}_d + \Sigma_s \bm{p} \cdot \bm{\gamma}_s,
\end{equation}
and each term is calculated by using the relation
\begin{gather}
\text{tr} \Sigma = 4 \Sigma_0 i\omega, \notag \\
\text{tr} (\gamma_d^i \Sigma) = 4 \Sigma_d p^i, \quad
\text{tr} (\gamma_s^i \Sigma) = 4 \Sigma_s p^i.
\end{gather}
The first equation leads to $\Sigma_0 = 0$, which is consistent with the Ward--Takahashi identity for the present model.
By introducing the spherical coordinate for momentum $\bm{k}$, we obtain
\begin{align}
\Sigma_d &= \frac{e^2}{(2\pi)^3 \varepsilon} v_d l \int \sin\theta d\theta d\phi \cos^2 \theta \frac{\sqrt{b}+c_d}{\sqrt{b}\sqrt{2a+2\sqrt{b}}}, \\
\Sigma_s &= \frac{e^2}{(2\pi)^3 \varepsilon} v_s l \int \sin\theta d\theta d\phi \cos^2 \theta \frac{\sqrt{b}+c_s}{\sqrt{b}\sqrt{2a+2\sqrt{b}}}.
\end{align}
The functions $a(\bm{k})$, $b(\bm{k})$, $c_d(\bm{k})$, and $c_s(\bm{k})$ are defined by
\begin{align*}
a(\bm{k}) &= (v_d^2 + v_s^2), \\
b(\bm{k}) &= (v_d^2 -v_s^2)^2 +3v_s^2 (4v_d^2-v_s^2) \frac{\tilde{k}^4}{k^4}, \\
c_d (\bm{k}) &= (v_d^2 +2v_s^2) \frac{k_x^2+k_y^2}{k^2} + (v_d^2-v_s^2) \frac{k_z^2}{k^2}, \\
c_s (\bm{k}) &= \frac{1}{2} (4v_d^2 -v_s^2) \frac{k_x^2 + k_y^2}{k^2} - (v_d^2-v_s^2) \frac{k_z^2}{k^2},
\end{align*}
with $\tilde{k}^4 = k_y^2 k_z^2 + k_z^2 k_x^2 + k_x^2 k_y^2$.
$\Sigma_d$ and $\Sigma_s$ give the beta functions for $v_d$ and $v_s$ as
\begin{equation}
\beta_{v_d} = \left.\frac{d\Sigma_d}{dl}\right|_{l=0}, \quad
\beta_{v_s} = \left.\frac{d\Sigma_s}{dl}\right|_{l=0}.
\end{equation}
These beta functions yield the RG equations for $v_d$ and $v_s$
\begin{align}
\label{eq:rg}
\frac{dv_d}{dl} = \beta_{v_d}, \quad \frac{dv_s}{dl} = \beta_{v_s}.
\end{align}
Note that when $v_s = 0$, the RG equations reduce to those for Dirac electrons in three dimensions, where we have $\beta_{v_d} = e^2/(6\pi^2 \varepsilon) \text{sgn}(v_d)$ and $\beta_{v_s} =0$ \cite{vishwanath}.

The set of RG equations \eqref{eq:rg} provides an RG flow on the $v_d$-$v_s$ plane [Fig.~\ref{fig:flow}(a)].
Both $v_d$ and $v_s$ become larger in low energies, and thus the point $v_d=v_s=0$ is unstable.
Indeed, the ratio of the two parameters $r\equiv v_s/v_d$ is important to determine the property of low-energy fixed points.
The RG equation for the ratio $r$ is obtained from Eq.~\eqref{eq:rg},
\begin{align}
\label{eq:beta_r}
\frac{dr}{dl} = \frac{\alpha}{2\sqrt{2}\pi^2} F(r),
\end{align}
where $\alpha \equiv e^2/(4\pi\varepsilon |v_d|)$ is a dimensionless coupling constant, and $F(r)$ is an odd function depending only on $r$.
The RG flow for the ratio $r$ is shown in Fig.~\ref{fig:flow}(b).
We can see two kinds of stable fixed points: One is at $r=v_s/v_d=0$, and the other at $r=\pm r_s$ with $r_s\approx 2.296$. The termination of a flow is determined solely by an initial ratio $r_0$, and does not depend on the absolute values of $v_d$ and $v_s$. The two types of stable fixed points are separated by unstable fixed points at $r=\pm 2$. The position of the fixed points corresponds to zeros of the function $F(r)$ [Fig.~\ref{fig:flow}(c)].
The properties of the fixed points are discussed after we see the renormalization of the coupling constant.

\begin{figure}
\centering
\includegraphics[width=0.7\hsize]{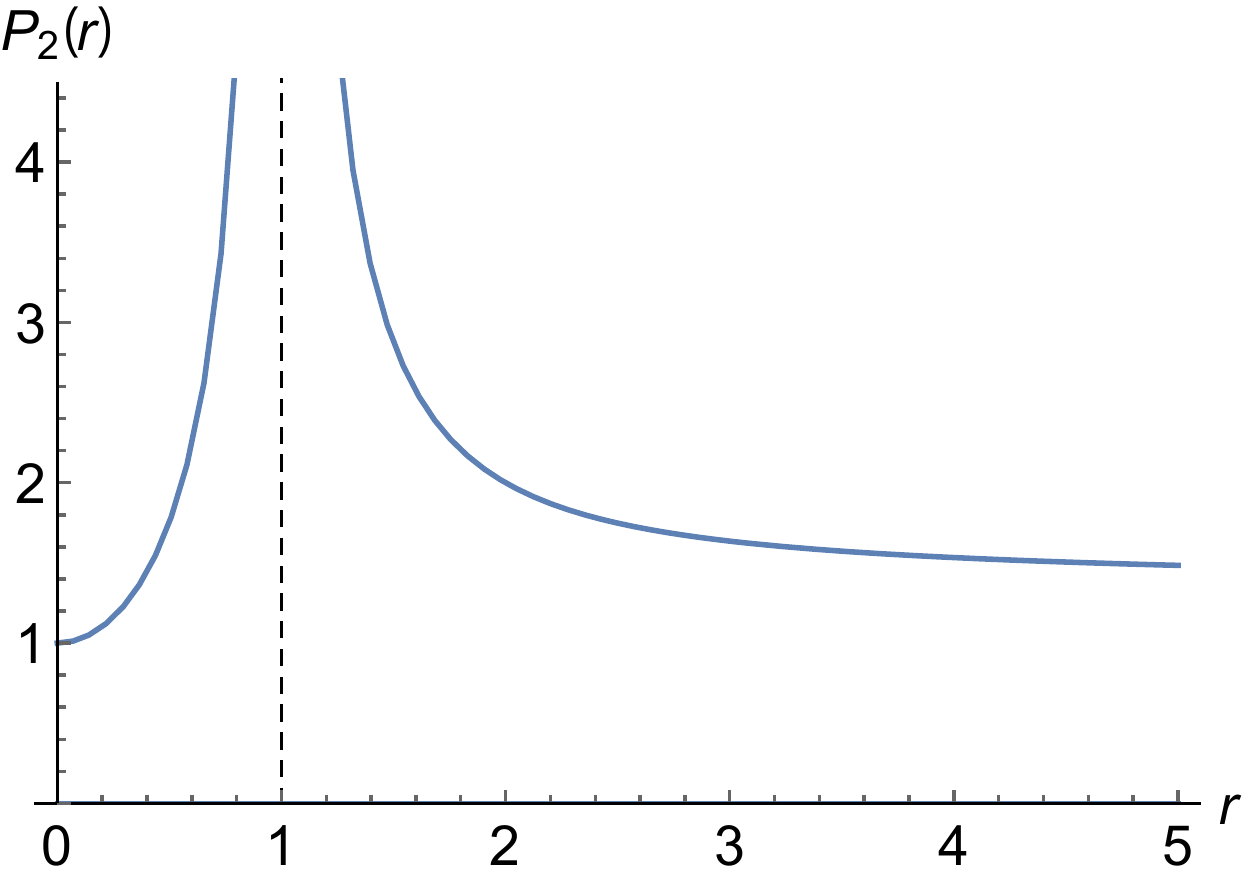}
\caption{
Function $P_2 (r)$ that characterizes the renormalization of the effective charge. It depends only on the ratio $r$. The singular behavior at $r=1$ comes from the line nodes.}
\label{fig:2}
\end{figure}

Next, we consider the one-loop polarization function $\Pi (\bm{q}, i\omega)$ [Fig.~\ref{fig:diagram}(b)], which yields the renormalization of the electric charge, given by
\begin{align}
\Pi (\bm{q}, i\omega) &= 2e^2 \int_{\bm{k},\omega'}' \text{tr} [G_0 (\bm{k}+\bm{q}, i\omega+i\omega') G_0 (\bm{k}, i\omega')] \notag \\
&= \Pi_2 q^2 + O(q^4),
\end{align}
where the factor 2 comes from a trace of $\tau$ matrices.
The polarization does not depend on the frequency $\omega$. When expanding it with respect to $q$, we can find a logarithmic divergence in the second-order term $\Pi_2$.
The divergence gives the renormalization of the electron charge, similarly to the self-energy considered above.
When we write $\Pi_2 = - e^2 q^2 l P_2(r)/(3\pi^2 v_d)$, the RG equation for the effective charge $g\equiv e/\sqrt{4\pi\varepsilon}$ is
\begin{equation}
\frac{dg^2}{dl} = - \frac{4g^4}{3\pi v_d} P_2(r).
\end{equation}
The even function $P_2(r)$ depends only on the ratio $r$ (Fig.~\ref{fig:2}).

For $r=0$, the system consists of four copies of isotropic Weyl fermions with $P_2(r)=1$, and together with Eq.~\eqref{eq:rg}, we can show that the dimensionless coupling constant $\alpha$ logarithmically decreases:  $\alpha(l) = \alpha_0 [1+(2\alpha_0/\pi) l]^{-1}$ \cite{vishwanath}.
For $r\neq 0$, $P_2(r) >0$ and the coupling constant also becomes weaker for lower energies, which justifies the perturbative RG treatment; the dimensionless coupling constant $\alpha$ has the unique stable fixed point at $\alpha=0$.
We observe the singularity at $r=1$, which originates from line nodes of the Fermi surface, elongating along the cubic axes.
This makes the density of states $D(E)\propto E$, in contrast to $D(E)\propto E^2$ for the case of the point node for $r\neq 1$, which changes the screening of charges. 
However, this is an artifact of the linearized theory, and the singularity arises only at $r=1$, so that it does not change the analysis of the fixed points.

\textit{Discussion}.
From the original Hamiltonian \eqref{eq:ham1} or \eqref{eq:ham2}, one would expect two fixed points: One is rotationally invariant ($v_1 \bm{k}\cdot\bm{J}$), and the other is Lorentz invariant ($v_d \bm{k}\cdot\bm{\gamma}_d$) \cite{note}.
Those two are indeed continuous symmetric points of the present model.
When a continuous symmetry is present, generators of the corresponding symmetry obey Lie algebra, i.e., the commutation relations must be closed.
Using this fact, we can identify symmetric points which have continuous symmetry.
For a linear combination of $\gamma_d^i$ and $\gamma_s^i$, the commutation relation is
\begin{equation}
[ a\gamma_d^i + b\gamma_s^i, a\gamma_d^j + b\gamma_s^j ]
= i \epsilon^{ijk} [ (-2a^2+b^2) \gamma_d^k + 2ab \gamma_s^k ].
\end{equation}
This has a closed form if and only if (1) $b = 0$ or (2) $b/a = \pm 2$.
Case (1) corresponds to $r=0$ ($v_s=0$), where the system is Lorentz invariant, and case (2) corresponds to $r = \pm 2$, which has rotational symmetry.
Otherwise, the model has no continuous symmetry, with at most the cubic symmetry $O_h$.

Since the RG flow is symmetric under the inversion of $r$, we concentrate our analysis on $r\geq 0$.
It is easily confirmed that the two symmetric points are fixed points, and actually we found the zeros of the function $F(r)$ at $r=0$ and $r=2$.
The question is whether they are stable or unstable.
Considering the symmetry of the model is controlled solely by the ratio $r$, we find that there is little likelihood of both points being stable.
Assuming that both are stable and that there is no other fixed point, $F(r)$ should touch but not cross zero at $r=2$. In this case, however, the point $r=2$ is subtle because it is stable for $r>2$ but unstable for $r<2$.

A more natural choice is that $F(r)$ crosses zero at $r=2$ to give other fixed points.
In other words, this system with seemingly two fixed points requires another fixed point for a reasonable RG flow.
From the one-loop RG analysis, we have observed in Figs.~\ref{fig:flow}(b) and \ref{fig:flow}(c) that the stable fixed point locates at $r=r_s(>2)$ and that hence $r=2$ is unstable.

In low energies, the system is either Lorentz or $O_h$ invariant.  The difference can be measured by angle-resolved photoemission spectroscopy, which directly observes the electron's energy band structure.  Another possible way of its detection is a measurement of magnetic susceptibility.  Because the system is isotropic (anisotropic) when it is Lorentz invariant ($O_h$ invariant), the measurement of the directional dependence of magnetic susceptibility may shed light on the electronic structure at low energies.

The important finding is that the $j=3/2$ Dirac fermions have the non-Lorentz-invariant stable fixed point in addition to the Lorentz-invariant fixed point.
The $O_h$-invariant stable fixed point appears because the two continuous symmetric points are not stable fixed points at the same time.
Restoration of the Lorentz invariance as a low-energy phenomenon is not universal when several continuous symmetries are present, and the property of a critical point will depend on the underlying symmetry of crystals. Further interesting physics topics may be hidden under this quantum criticality.

\textit{Acknowledgments}.
We thank T. H. Hsieh and E.-G. Moon for useful comments. 
This work was supported in part by the MRSEC Program of the National Science Foundation under award No.~DMR-1419807.


\begin{thebibliography}{99}
\bibitem{note}
The Hamiltonian $H=v_d \bm{k}\cdot\bm{\gamma}_d$ itself is not Lorentz invariant in a sense that $v_d$ is not equal to the speed of light $c$ in the material. In this case, considering the relativistic effect in RG analysis restores the Lorentz invariance in the low-energy limit \cite{rg1}.
\end{thebibliography}
\end{document}